\documentclass[twocolumn,showpacs,preprintnumbers,amsmath,amssymb,prl,superscriptaddress]{revtex4-1}
\usepackage{bbm}
\usepackage{mathrsfs}
\usepackage{graphicx}
\usepackage{dcolumn}
\usepackage{bm}
\usepackage{amsmath}
\usepackage{amsfonts}
\usepackage{color}
\usepackage[colorlinks=true,citecolor=blue,anchorcolor=blue]{hyperref}
\usepackage{floatrow}

\begin{document}

\title{Multiple Chiral Majorana Fermion Modes and Quantum Transport}
\author{Jing Wang}
\thanks{wjingphys@fudan.edu.cn}
\affiliation{State Key Laboratory of Surface Physics, Department of Physics, Fudan University, Shanghai 200433, China}
\affiliation{Collaborative Innovation Center of Advanced Microstructures, Nanjing 210093, China}
\author{Biao Lian}
\thanks{biao@princeton.edu}
\affiliation{Princeton Center for Theoretical Science, Princeton University, Princeton, New Jersey 08544, USA}

\begin{abstract}
We propose a general recipe for chiral topological superconductor (TSC) in two dimensions with multiple $N$ chiral Majorana fermion modes from a quantied anomalous Hall insulator in proximity to an $s$-wave superconductor with nontrivial band topology. A concrete example is that a $N=3$ chiral TSC may be realized by coupling a magnetic topological insulator and the ion-based superconductor such as FeTe$_{1-x}$Se$_{x}$ ($x=0.45$). We further propose the electrical and thermal transport experiments to detect the Majorana nature of three chiral edge fermions. A unique signature is that the two-terminal electrical conductance of a quantized anomalous Hall-TSC junction obeys a distribution averaged to $(2/3)e^2/h$, which is due to the random edge mode mixing of chiral Majorana fermions and is distinguished from possible trivial explanations.
\end{abstract}

\date{\today}


\maketitle

Majorana fermions have attracted intense interest in both particle physics and condensed matter physics~\cite{wilczek2009,elliott2015}. The chiral Majorana fermion, a massless fermionic particle being its own antiparticle, could arise as a one-dimensional (1D) quasiparticle edge state of a 2D topological states of quantum matter~\cite{moore1991,read2000,mackenzie2003,kitaev2006,fu2008,sau2010,alicea2010,qi2009}. The propagating chiral Majorana fermions could lead to non-abelian braiding~\cite{lian2017} and may be useful in topological quantum computation~\cite{kitaev2003,nayak2008}. A simple example hosting chiral Majorana fermion mode (CMFM) is the $p_x+ip_y$ chiral topological superconductor (TSC) with a Bogoliubov-de Gennes (BdG) Chern number $N=1$, which can be realized from a quantum anomalous Hall (QAH) insulator with a Chern number $C=1$ in proximity to a conventional $s$-wave superconductor~\cite{qi2010b,chung2011,wang2015c}. The quantum transport in a QAH-TSC-QAH (QTQ) junction is predicted to exhibit a half quantized conductance plateau induced by a single CMFM~\cite{chung2011,wang2015c,lian2016a}, which has been recently observed in Cr$_x$(Bi,Sb)$_{2-x}$Te$_3$ (CBST) thin film QAH system in proximity with Nb superconductor~\cite{he2017}.

Physically, the $N=1$ chiral TSC emerges in the neighborhood of the QAH plateau transitions, where the superconducting pairing gap exceeds QAH gap~\cite{qi2010b}, and it can be driven by an external magnetic field or electric field in magnetic topological insulators (MTIs)~\cite{wang2014a,kou2015,fengy2015,wang2016b}. The magnetic field at coercivity inevitably introduces random domains, making MTIs to be strongly disorderd~\cite{yasuda2017b}. The single CMFM in this system is robust against disorder~\cite{lian2018}. However, alternative explanations of the half plateau without CMFM under strong disorders have been proposed~\cite{ji2018,huang2018}, which arises from incoherence due to disorder. The noise and interferences measurement may distinguish chiral Majorana fermion from the disorder-induced metallic phases~\cite{chung2011,strubi2011,li2018, fu2009a,akhmerov2009,law2009}  . 

In this Letter, we propose a general recipe for a higher odd Chern number $N$ chiral TSC which supports multiple CMFMs. The random edge mode mixing of chiral Majorana fermions lead to novel quantized transport. In sharp contrast to the previous proposal that chiral TSC is achieved near the QAH plateau transition~\cite{qi2010b,wang2015c}, where strong disorders accompany in the system. Here the system we proposed is homogenous, which provides an ideal platform for studying the exotic physics of chiral Majorana fermions.

\begin{figure}[b]
\begin{center}
\includegraphics[width=1.6in]{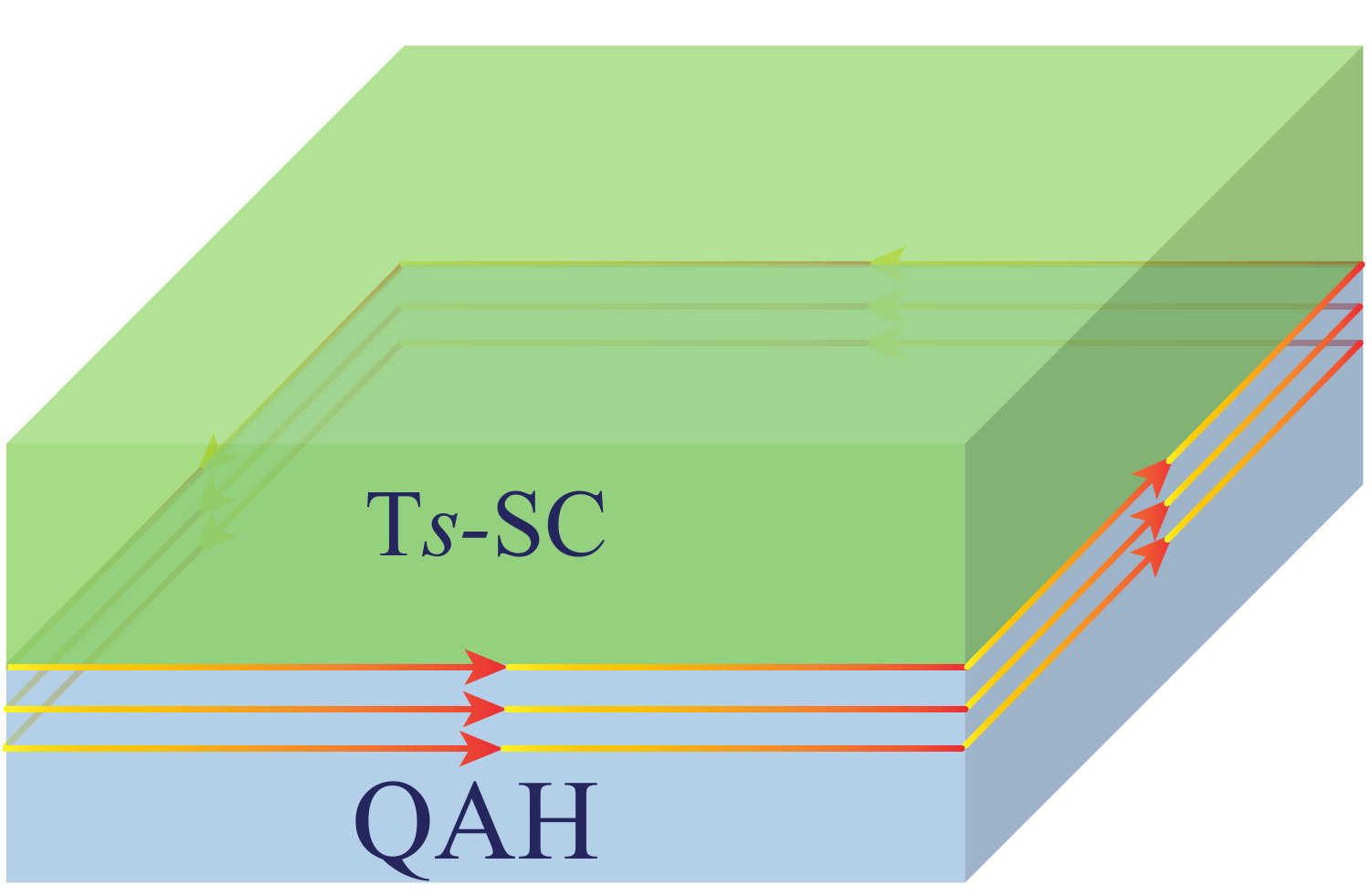}
\end{center}
\caption{The heterostructure for chiral TSC with an odd number of CMFMs consists of QAH in a MTI and a T$s$-SC on top. Take $C=1$ QAH for exmaple, a $N=3$ TSC is realized when the exchange field is large enough. When QAH has a higher Chern number, a higher odd number $N$ TSC may be realized.} \label{fig1}
\end{figure}

\emph{Model.}
The basic mechanism for 2D chiral TSC is to introduce $s$-wave superconductivity ($s$-SC) and ferromagnetism (FM) into a strong spin-orbit coupled (SOC) system, such as the spin-helical surface states (SSs) of TIs~\cite{hasan2010,qi2011}. Instead of inducing superconductivity into a MTI for chiral TSC, one can introduce the FM proximity effect into superconducting Dirac SSs, where the CMFM exists at the boundary between FM and superconductor~\cite{fu2009a}. The latter one is more practical for a homogenous system, since FM exchange coupling is usually much larger than conventional $s$-SC proximity. Therefore, it is natural to ask whether exotic topological states exist in the heterostructure of topological FM insulator and $s$-SC with nontrivial band topology dubbed as topological $s$-SC (T$s$-SC) shown in Fig.~\ref{fig1}. The T$s$-SC has a fully bulk pairing gap and $s$-SC gap on the single spin-helical Dirac SS. The prototype T$s$-SC materials are the ion-based superconductors such as FeTe$_{0.55}$Se$_{0.45}$ (FTS)~\cite{zhang2018}. The general theory presented here for chiral TSC is generic for the higher Chern number QAH insulator~\cite{wang2013a} and T$s$-SC. We would like to start with a simple model describing the $C=1$ QAH in MTIs~\cite{chang2013b} for concreteness. The low energy physics of the heterostructure is described by four Dirac SSs only, for the bulk states in MTI and T$s$-SC are gapped. The generic form of the 2D effective Hamiltonian without superconducting proximity effect is
\begin{equation}
\begin{aligned}
\mathcal{H}(\mathbf{k}) &=\begin{pmatrix}
H_1 & V\\
V^\dag & H_2
\end{pmatrix}.
\end{aligned}
\end{equation}
Here $H_1$ describes the T$s$-SC SSs with proximity effect from MTI, the bulk metallic states in T$s$-SC are neglected since they are gapped with superconducting pairing, $H_2$ describes the QAH in MTI film,
\begin{equation}
\begin{aligned}
H_1 & = v_1k_y\sigma_1\tau_3-v_1k_x\sigma_2\tau_3+\frac{\lambda_1}{2}\sigma_3(1-\tau_3)+2\delta,\\
H_2 & = v_2k_y\sigma_1\tau_3-v_2k_x\sigma_2\tau_3+m(k)\tau_1+\lambda_2\sigma_3,
\end{aligned}
\end{equation}
with the basis of $\varphi^i_{\mathbf{k}}=(c_{t_i\uparrow},c_{t_i\downarrow},c_{b_i\uparrow},c_{b_i\downarrow})^T$, ($i=1,2$), where $t$ and $b$ denote the top and bottom SSs and $\uparrow$ and $\downarrow$ represent the spin up and down states, respectively. $\sigma_j$ and $\tau_j$ ($j=1,2,3$) are Pauli matrices acting on spin and layer, respectively.  $v_i$ is the Fermi velocity, which have opposite signs in FTS and MTI~\cite{zhang2018,wangzj2015,wu2016}. (The relative sign of velocities doe not affect the results). $\lambda_i$ is the FM exchange field along the $z$ axis which can be tuned by a magnetic field. The short-range FM proximity effect only affects the bottom SS of T$s$-SC and $\lambda_1\leq\lambda_2$. $m(k)=m_0+m_1|\mathbf{k}|^2$ is the hybridization between top and bottom SSs in MTI. $|\lambda_2|>|m_0|$ guarantees QAH state in MTI. $2\delta$ is the energy band alignment between two Dirac cones. For simplicity, we set $v_2=-v_1\equiv v$, 
$\lambda_1=\lambda_2\equiv\lambda$, and neglect the inversion symmetry breaking in each material. $V=g\tau_-$ is the hybridization between the bottom T$s$-SC and top MTI surfaces at interface, where $\tau_-=(\tau_1-i\tau_2)/2$, $g$ is real constant.

With superconducting proximity effect, a finite pairing amplitude is induced in MTI and T$s$-SC SSs. The BdG Hamiltonian becomes $H_{\text{BdG}}=(1/2)\sum_{\mathbf{k}}\Psi^\dag_\mathbf{k}\mathcal{H}_{\text{BdG}}(\mathbf{k})\Psi_{\mathbf{k}}$, with $\Psi_\mathbf{k}=(\psi^T_\mathbf{k},\psi^\dag_{-\mathbf{k}})^T$, $\psi_{\mathbf{k}}=(\varphi^{1}_{\mathbf{k}},\varphi^2_{\mathbf{k}})$ and
\begin{equation}
\begin{aligned}
\mathcal{H}_{\text{BdG}}(\mathbf{k}) &=
\begin{pmatrix}
\mathcal{H}(\mathbf{k})-\mu & \Delta(\mathbf{k})\\
\Delta^\dag(\mathbf{k}) & -\mathcal{H}^*(-\mathbf{k})+\mu
\end{pmatrix},\label{BdG}
\\
\Delta(\mathbf{k}) &=\begin{pmatrix}
\Delta_1(\mathbf{k}) & 0\\
0 & \Delta_2(\mathbf{k})
\end{pmatrix},
\end{aligned}
\end{equation}
Here $\mu$ is the chemical potential relative to the Dirac cone in MTI, $\Delta_1(\mathbf{k})=i\Delta_1\sigma_2$ and $\Delta_2(\mathbf{k})=i(\Delta_2^t/2)\sigma_2(1+\zeta_3)+i(\Delta_2^b/2)\sigma_2(1-\zeta_3)$ with $\zeta_3$ the Pauli matrix in Nambu space. $\Delta_1$, $\Delta_2^t$, and $\Delta_2^b$ are pairing gap function in SSs of T$s$-SC, top, and bottom MTI. All $\Delta_i$ are chosen as $\mathbf{k}$ independent, since they are induced by the $s$-SC proximity effect, for example from the bulk hole pocket at $\Gamma$ point in FTS. Usually $\Delta_1\geq\Delta_2^t\gg\Delta_2^b$. Here we set $\Delta_1=\Delta_2^t\equiv\Delta$ and $\Delta_2^b=0$, which is realistic in superconducting proximity effect between Bi$_2$Te$_3$ thin film and FTS with short coherence length~\cite{chenm2018}.

\begin{figure}[b]
\begin{center}
\includegraphics[width=3.3in]{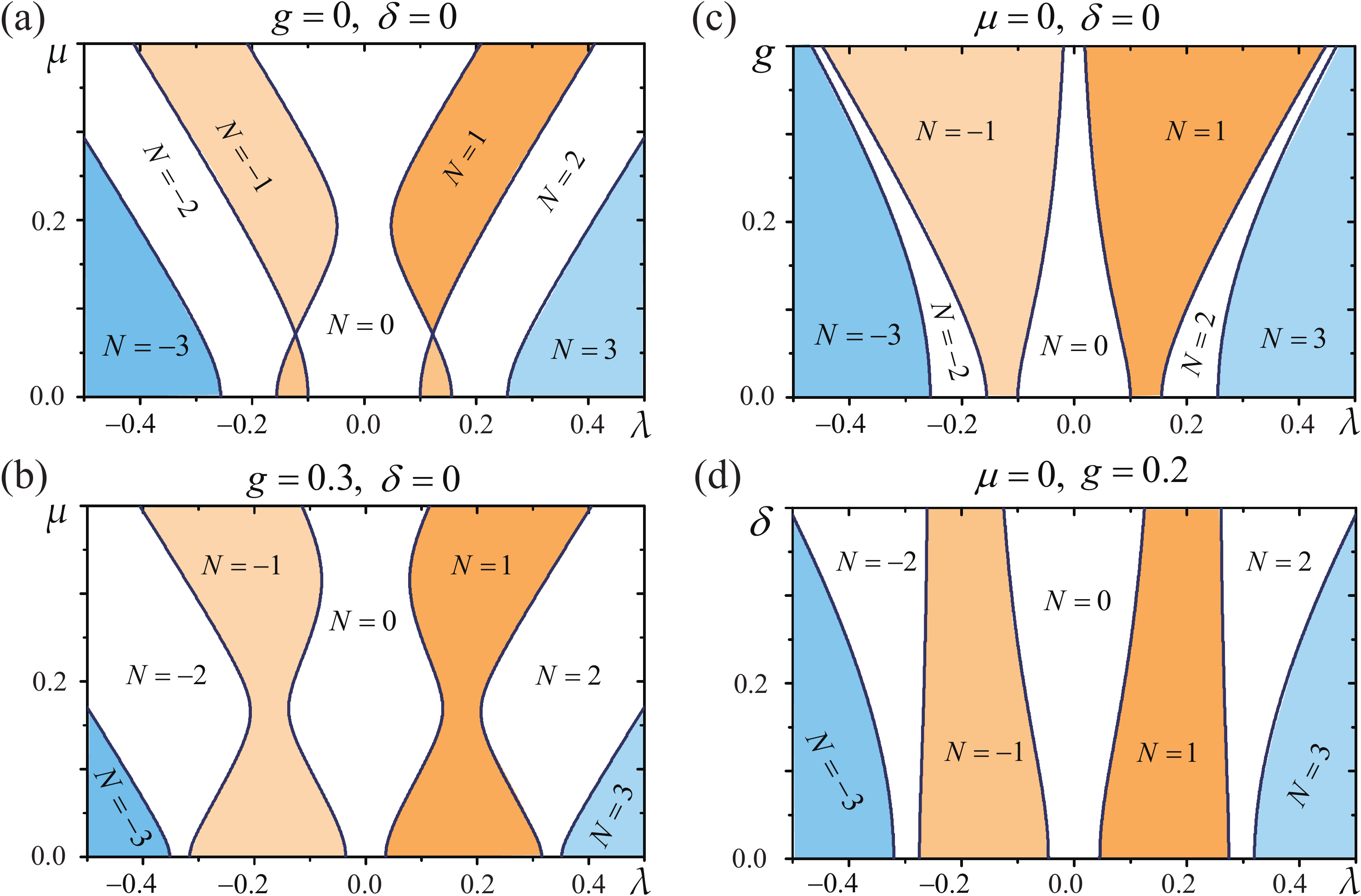}
\end{center}
\caption{Phase diagram of the heterostructure with typical parameters. (a) $g=0$, $\delta=0$. (b) $g=0.3$, $\delta=0$. (c) $\mu=\delta=0$. The even $N=0,2$ phases disappear when $m_0=0$ and the phase boundary between $N=1$ and $N=3$ is $g=\sqrt{\lambda^2-\Delta^2}$. (d) $\mu=0$, $\delta=-0.1$. All other parameters $v=1$, $\Delta=0.1$, $m_0=0.2$.}
\label{fig2}
\end{figure}

\emph{Phase diagram.} 
The BdG Hamiltonian in Eq.~(\ref{BdG}) can be classified by the Chern number $N$. Since the topological invariants cannot change without closing the bulk gap, the phase diagram can be determined by first finding the phase boundaries as gapless regions in parameter spaces, and then calculate $N$ of the gapped phases. 
To start, we first consider the phase diagram in the limit $g=0$, in which case the system is decoupled into two BdG models $H^{\text{BdG}}_1$ and $H^{\text{BdG}}_2$,
\begin{equation}
H_i^{\text{BdG}} =
\begin{pmatrix}
H_i(\mathbf{k})-\mu & \Delta_i(\mathbf{k})\\
\Delta_i^\dag(\mathbf{k}) & -H_i^*(-\mathbf{k})+\mu
\end{pmatrix}.
\end{equation}
Here $H_1^{\text{BdG}}$ is the superconducting Dirac SSs of T$s$-SC with only the bottom SS in proximity to FM. The top and bottom SSs in T$s$-SC are further decoupled. The energy spectrum of the top SS is $E^{1,t}_{\mathbf{k}}=\pm\sqrt{(\pm v\left|\mathbf{k}\right|-\mu')^2+\Delta^2}$, and $\mu'=\mu-2\delta$, which resembles that of the spinless $p_x+ip_y$ superconductor but respects time-reversal symmetry~\cite{fu2008}. The excitation spectrum of the bottom SS is $E^{1,b}_{\mathbf{k}}=\pm\sqrt{\Delta^2+\lambda^2+\mu'^2+v^2|\mathbf{k}|^2\pm\sqrt{\lambda^2(\Delta^2+\mu'^2)+\mu'^2v^2|\mathbf{k}|^2}}$, with the gap closing point at $\mathbf{k}=0$ and $\lambda=\sqrt{\Delta^2+\mu'^2}$. For $\left|\lambda\right|<\sqrt{\Delta^2+\mu'^2}$, the bottom SS is adiabatically connect to the top SS in the $\lambda=0$ limit, so they are topologically equivalent. Therefore, the whole T$s$-SC SS possesses non-trivial topology, but there is no chiral edge state, since there is no geometric edge to the 2D surface of a 3D bulk. For $\left|\lambda\right|>\sqrt{\Delta^2+\mu'^2}$, the bottom SS is adiabatically connected to FM with $\Delta=\mu'=0$ which is topologically trivial, so there exists a single CMFM at the edge domain boundary at T$s$-SC bottom, and $N_1=\text{sgn}(\lambda)$ is the sign of $\lambda$. $H_2^{\text{BdG}}$ is the superconducting proximity coupled QAH insulator, which has been studied in Ref.~\cite{wang2015c}. For $\mu=0$, $N_2=0$ for $|\lambda|<\lambda_c^-$ (which vanishes when $m_0=0$), $N_2=\text{sgn}(\lambda)$ for $\lambda_c^-<|\lambda|<\lambda_c^+$, and $N_2=2\text{sgn}(\lambda)$ for $|\lambda|>\lambda_c^+$. Here $\lambda_c^{\pm}=(\sqrt{4m_0^2+\Delta^2}\pm\Delta)/2$. A finite $\mu$ enlarges the odd $N_2$ TSC phase. The total Chern number of the heterostructure without $V$ is $N=N_1+N_2$. The phase diagram with parameters $(\mu,\lambda)$ is shown in Fig.~\ref{fig2}(a), where the different chiral TSC phases are denoted by the corresponding Chern numbers. 

Next, we study the effect of $V$ at interface. Similar to the $g=0$ case, we determine the phase boundaries by the bulk gap closing regions in Eq.~(\ref{BdG}), which is always at $\mathbf{k}=0$ point. As  show in Fig.~\ref{fig2}(b), when the $g$ term is turned on, it makes the chiral TSC phase with the same Chern numbers simply connected. Meanwhile, it shrinks the $N=0$ phase and enlarges the $N=1$ phase, and further pushes the phase boundary between $N=2$ and $N=3$ towards a larger $\lambda$. For a given exchange field, $\mu$ will drive the system into TSC phases with smaller $N$. Therefore, one optimal condition for $N=3$ TSC is $\mu=0$, which corresponds to undoped QAH system. This is just the opposite to the optimal condition $\mu\neq0$ for obtaining the $N=1$ TSC phase from the QAH plateau transition~\cite{wang2015c}. As shown in Fig.~\ref{fig2}(c), $g$ enlarges the $N=1$ phase and shrinks all other $N$ phases. This can be understood from the band crossing at the interface. The single-particle Hamiltonian at interface is  $H_{\text{int}} = vk_y\sigma_1-vk_x\sigma_2+\lambda\sigma_3+g\tau_1+\delta(1+\tau_3)$, with the energy spectrum $E_{\text{int}}=\delta\pm(\sqrt{g^2+\delta^2}\pm\sqrt{\lambda^2+v|\mathbf{k}|^2})$. $g$ splits the two copies of Dirac bands up and down in energy. Whenever the Dirac band edge crosses the chemical potential, $N$ reduces by one. As shown in Fig.~\ref{fig2}(d), $\delta$ enlarges the trivial even $N$ TSC, and shrink the nontrivial odd $N$ TSC towards larger $\lambda$. Similarly, $m_0$ enlarges $N=0$ TSC and shrinks $N\neq0$ TSC. Thus $m_0=\delta=0$ is preferred for higher $N$ TSC. For a simple case $m_0=\delta=0$ and infinitesemal $\Delta$, a simple sum rule for Chern number of the heterostucture is
\begin{equation}
N=N_2+\text{sgn}\left(|\lambda|-|g|\right)N_1.
\end{equation}
In general, the coupling $g$ will strongly modify the Chern number of the heterostructure from that of the decoupled systems. The  chiral TSC with higher odd Chern numbers requires a large enough exchange field, and is simply obtained by growing multilayer heterostructure or using higher Chern number QAH following the above recipe.

\begin{figure}[b]
\begin{center}
\includegraphics[width=2.5in]{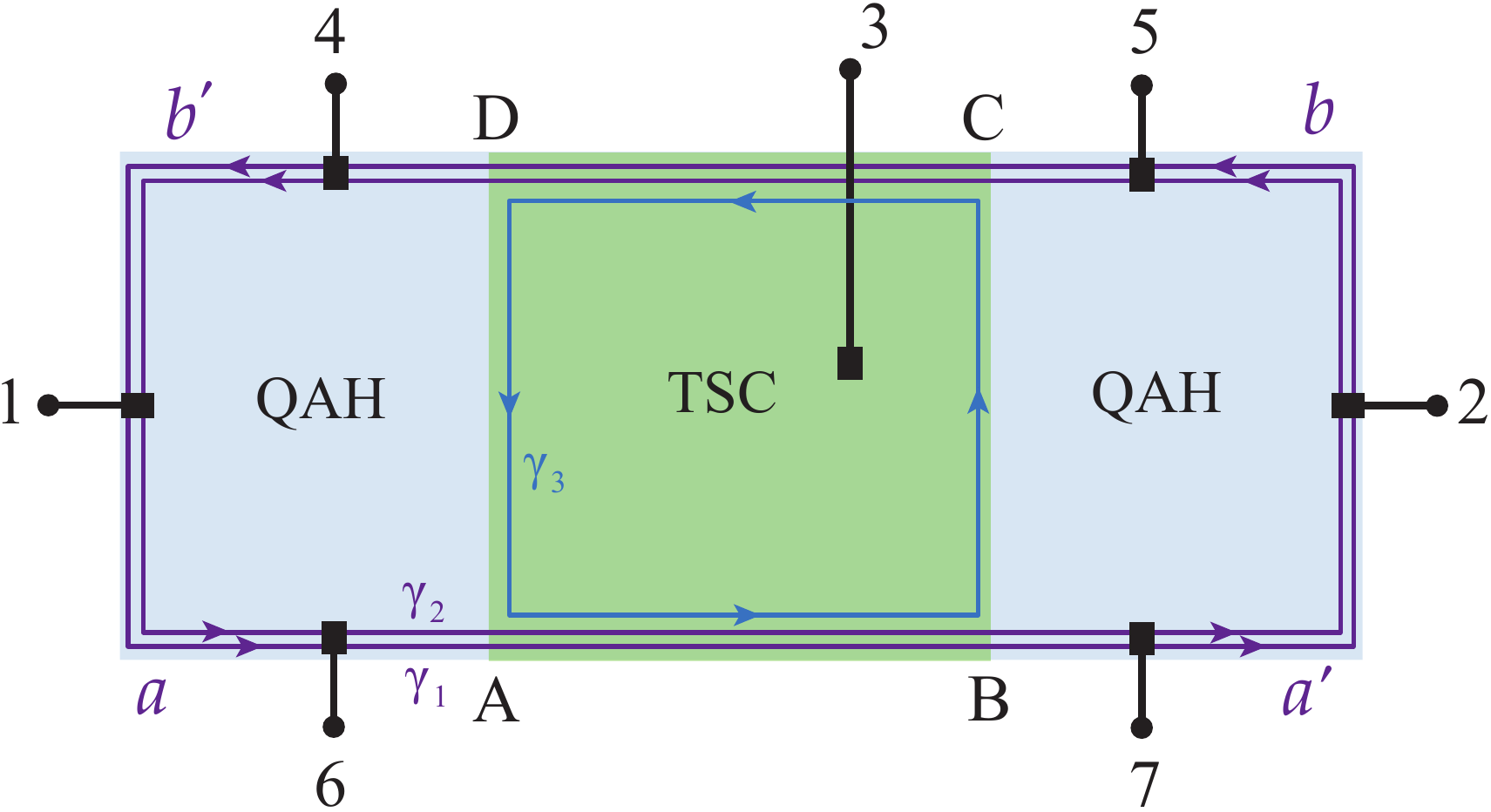}
\end{center}
\caption{The transport configuration of a QTQ ($N$-$N'$-$N$=$2$-$3$-$2$) junction device. The arrows on edge represents CMFMs.}
\label{fig3}
\end{figure}

\emph{Transport.} 
To probe the multiple neutral CMFMs, we consider the electrical and thermal transports in $N=3$ chiral TSC. The Hall bar device we shall disucss is a QTQ junction as shown in Fig.~\ref{fig3}, which has been studied for $N=1$ and $N=2$ chiral TSCs. Both the left and right QAH regions have Chern number $C=1$, and thus have a charged chiral fermion mode on their edges with vacuum. The charge chiral fermion mode can be equivalently written as two CMFMs $\gamma_1$ and $\gamma_2$ as shown in Fig. \ref{fig3}, and the electron annihilation operators $a,a',b,b'$ on the left (right) bottom (top) QAH edges are locally related to the CMFMs as $a,a',b,b'=\gamma_1+i\gamma_2$. There exists a third CMFM $\gamma_3$ on the vertical edges between $C=1$ QAH and $N=3$ TSC, which merges with $\gamma_1$ and $\gamma_2$ on the top and bottom TSC edges.

We shall assume the electrical current is only applied at terminals 1, 2 and 3, while terminals 4 to 7 are only used as voltage leads. Lead on electrode 3 is connected to the TSC bulk, while all the other leads are on the edge. The electrical transport of the superconducting junction is governed by the generalized Landauer-B\"uttiker formula \cite{anantram1996,entin2008,chung2011,wang2015c}, which takes the form among leads 1-3 as
\begin{equation}\label{LB}
\begin{aligned}
I_1 &= \frac{e^2}{h}[(1-r+r^A)(V_1-V_{s})+(t-t^A)(V_{s}-V_2)],
\\
I_2 &= \frac{e^2}{h}[(1-r+r^A)(V_2-V_{s})+(t-t^A)(V_{s}-V_1)],
\\
I_3 &= -I_1-I_2,\quad V_3=V_{s},
\end{aligned}
\end{equation}
where $I_i$ and $V_i$ are the inflow current and voltage of lead $i$, $V_{s}$ is the voltage of the TSC, and we have assumed the contact resistance vanishes between lead $3$ and the TSC bulk, which is appropriate when the electrodes are good metals. Here $r,r^A,t$ and $t^A$ are the normal reflection, Andreev reflection, normal transmission and Andreev transmission probabilities between leads $1$ and $2$, respectively, which satisfy $r+r^A+t+t^A=1$.

To examine the normal and Andreev probabilities, consider the charged chiral fermion mode $a=\gamma_1+i\gamma_2$ incident from lead $1$. When propagating on the bottom TSC edge A-B, it could randomly mix with $\gamma_3$ due to unavoidable edge disorders. Therefore, when the incident charge mode $a$ reaches corner B, it has the normal and Andreev probabilities $t_{1}$ and $t^A_{1}$ to become $a'$ and $a'^\dag$, but also has a remaining probability $p^{(1)}=1-t_{1}-t^A_{1}$ to propagate as $\gamma_3$. The $\gamma_3$ mode will then circulate along the TSC edge, and has a propagation probability into charge modes $b'$, $b'^\dag$ (or $a'$, $a'^\dag$) whenever it reaches corner D (or B), thus contributing probabilities $r_{n}$ and $r^A_{n}$ (or $t_{n}$ and $t^A_{n}$) during its $n$-th lap. Summing over $n$ then yields the total $r,r^A,t$ and $t^A$. Such a summation is difficult. However, since $\gamma_3$ is charge neutral, its propagation probability into electron and hole states will always be equal, so we conclude $r_{n}=r^A_{n}$ for any $n$, and $t_{n}=t^A_{n}$ for all $n\ge2$. Therefore, we find $t-t^A=t_{1}-t^A_{1}$, and $r-r^A=0$, which are the only quantities needed in the Landauer-B\"uttiker formula of Eq.~(\ref{LB}).

Next we calculate $t_{1}-t^A_{1}$. The chemical potential, hopping and pairing on TSC edge A-B yields a term $H_R=i\gamma^TL(x)\gamma$ in the Hamiltonian, where $\gamma=(\gamma_1,\gamma_2,\gamma_3)^T$ is the CMFM basis, and $L(x)$ is a $3\times3$ real antisymmetric matrix. In terms of the SO(3) group generators $\mathbf{T}=(T_1,T_2,T_3)$, one can rewrite it as $L=i\omega\mathbf{n}\cdot\mathbf{T}$, where $|\mathbf{n}|=1$. For a given edge, one expects $\omega=\langle\omega\rangle+\delta\omega(x)$ and $\mathbf{n}=\langle \mathbf{n}\rangle +\delta \mathbf{n}(x)$ to fluctuate, where the fluctuations are usually small compared to the mean values, i.e., $|\delta\mathbf{n}(x)|\ll|\langle \mathbf{n}\rangle|$. This leads to a SO(3) transformation $Q\approx e^{i\phi\langle \mathbf{n}\rangle\cdot\mathbf{T}}$ of $\gamma$, where $\phi\approx\int_A^B\omega(x)dx/v_M$ is uniformly random in $[0,2\pi)$ when edge A-B is long enough, $v_M$ is the average Majorana velocity~\cite{levin2007,lee2007,lian2018a}. The average normal and Andreev transmissions along the edge are thus the mean value over $\phi$:
\begin{equation}\label{ttA}
t_{1}=\overline{|u^\dag Q u|^2},\quad t^A_{1}=\overline{|u^T Qu|^2},
\end{equation}
where $u=(1,i,0)^T/\sqrt{2}$ is the electron annihilation operator under Majorana basis. The result gives $t_1-t_1^A=\cos^2\theta$, where $\theta$ is the angle between $\langle \mathbf{n}\rangle$ and the $\gamma_3$ axis (see the Supplementary Material~\cite{supple}). By defining $\sigma_{12}=I/(V_1-V_2)$ for current $I=I_1=-I_2$ applied between leads $1$ and $2$ ($I_3=0$), and $\sigma_{13}=I/(V_1-V_3)$ for current $I=I_1=-I_3$ applied at leads $1$ and $3$ ($I_2=0$), we obtain $\sigma_{12}=(1+\cos^2\theta)/2$ and $\sigma_{13}=1-\cos^4\theta$ in units of $e^2/h$. Note that $\langle\mathbf{n}\rangle$ depends on samples and physical conditions, so if we assume $\langle\mathbf{n}\rangle$ distributes uniformly on the unit sphere $S^2$, we obtain the probability distributions of $\sigma_{12}$ and $\sigma_{13}$ among various samples or physical conditions
\begin{equation}\label{sigma1213}
p(\sigma_{12})=\frac{1}{\sqrt{2\sigma_{12}-1}}, \quad p(\sigma_{13})=\frac{1}{4(1-\sigma_{13})^{3/4}},
\end{equation}
where $\sigma_{12}\in[\frac{1}{2},1]$ and $\sigma_{13}\in[0,1]$ in units of $e^2/h$. The average values of $\sigma_{12}$ and $\sigma_{13}$ can then be derived to be $\overline{\sigma}_{12}=(2/3)e^2/h$ and $\overline{\sigma}_{13}=(4/5)e^2/h$. Moreover, if the TSC edge is in the strong fluctuation limit $|\delta\mathbf{n}(x)|\gg|\langle \mathbf{n}\rangle|$ (which is less likely), we would have $t=t^A$, leading to quantized conductances $\sigma_{12}=e^2/2h$ and $\sigma_{13}=e^2/h$~\cite{supple}.

In addition, one can derive the resistance matrix measured from other leads are $R_{12,46}=R_{12,57}=h/e^2$, and $R_{12,45}=R_{12,67}=\sigma_{12}^{-1}-h/e^2$, where $R_{ij,kl}\equiv (V_k-V_l)/I$ with current $I$ applied between leads $i$ and $j$~\cite{supple}.

\begin{figure}[t]
\begin{center}
\includegraphics[width=3.3in]{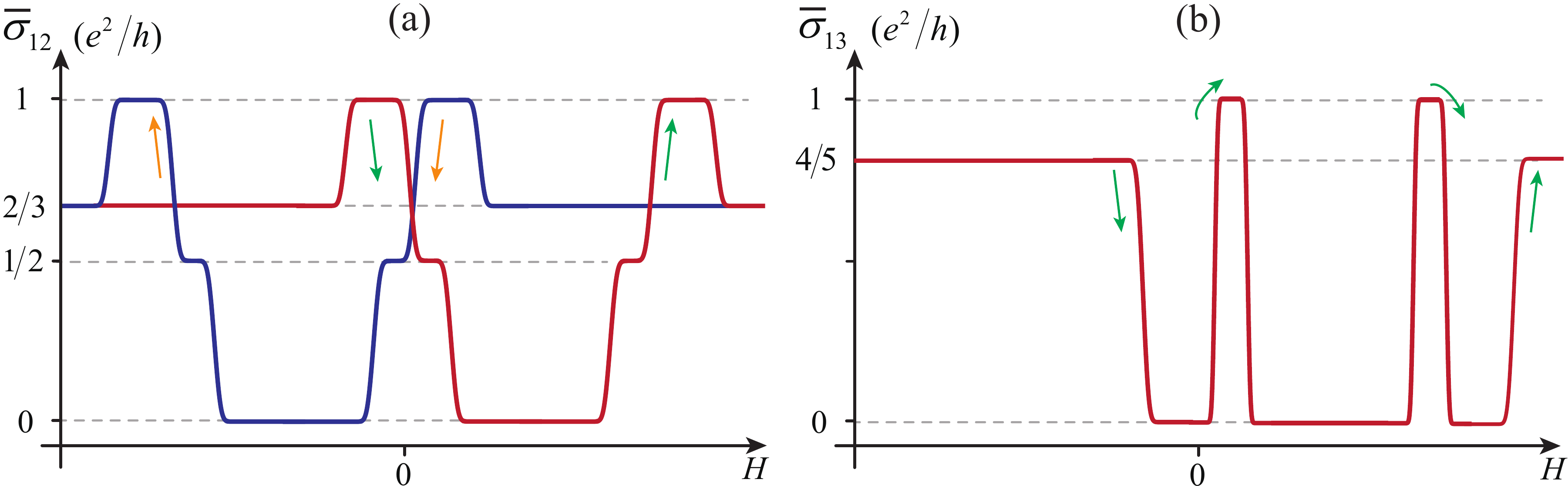}
\end{center}
\caption{(a) The average value $\overline{\sigma}_{12}$ generically shows plateau transition $2/3\rightarrow1\rightarrow1/2\rightarrow0\rightarrow1/2\rightarrow1\rightarrow2/3$ in unit of $e^2/h$ during the hysteresis loop. (b) The average value $\overline{\sigma}_{13}$ shows $4/5$ and $1$ peaks for $|N|=3$ and $|N|=1$ TSC phases, respectively. Here only one cycle of hysteresis loop is shown.}
\label{fig4}
\end{figure}

The exchange field can be tuned by either a perpendicular or an in-plane external magnetic field. Therefore, the TSC phases will experience the BdG Chern number variation $N=3\rightarrow2\rightarrow1\rightarrow0\rightarrow-1\rightarrow-2\rightarrow-3$ as $\lambda$ decreases in the hysteresis loop. Meanwhile, the QAH phase will experience the Chern number change $C=1\rightarrow0\rightarrow-1$, and in terms of $N=2\rightarrow0\rightarrow-2$. In general, the average $\overline{\sigma}_{12}$ will exhibit the plateau transition as shown in Fig.~\ref{fig4}(a). Since the system in the magnetized state without external magnetic fields is homogenous in the sense of weak disorder without percolation transition, the unique $2/3$ quantized average conductance plateau manifests the $N=3$ TSC. 

Finally, we discuss the thermal transport. The $N=3$ chiral TSC exhibits a quantized thermal Hall conductance $\kappa_{xy}=3$ in units of $\kappa_0=(\pi^2/6)(k_B^2/h)T$, where $k_B$ is the Boltzmann constant and $T$ is temperature. Moreover, the QTQ junction will exhibit quantized thermal resistances resembling the electric resistances of a filling factor 2-3-2 integer quantum Hall junction~\cite{williams2007,kim2007}. For a heat current applied between leads $1$ and $2$, the thermal resistances are given by $R^Q_{12,46}=R^Q_{12,57}=1/2$, and $R^Q_{12,45}=R^Q_{12,67}=1/6$ in unit of $1/\kappa_0$. Generically, phonons and magnons also contribute to the thermal conductance, which will deviate from the quantized value. However, their contribution can be well distinguished from the temperature dependence \cite{banerjee2017}.

\emph{Discussion.}  We discuss the experimental feasibility of higher odd Chern number chiral TSC. The key point is to invert the bands by a large exchange field, while keeping the QAH insulating. The hybridization between top and bottom SSs in QAH better to be small. For QAH in magnetic TIs, the exchange field $\lambda\approx30$~meV in CBST~\cite{lee2015}, and is $50$~meV in V-I codoped TI~\cite{qi2016}. $m_0$ vanishes when film thickness exceeds five quintuple layers. For T$s$-SC in FTS, $\Delta=2$~meV and $\mu=5$~meV below $T_c=14.5$~K~\cite{zhang2018}. The work function in FTS grown on SrTiO$_3$ is around $4.35\pm0.1$~eV, which is in the same range for that in (Bi,Sb)$_2$Te$_3$ thin film on SrTiO$_3$ about $4.4$~eV. Therefore, $\delta$ can be tuned to be small. $g$ is unknown, but can be tuned by inserting an insulating ultrathin layer between T$s$-SC and QAH. Other possible T$s$-SC materials include ion-based superconductor such as BaFe$_2$As$_2$, LiFeAs~\cite{zhang2018b}, and the superconducting doped TIs such as Cu$_x$Bi$_2$Se$_3$, Tl$_x$Bi$_2$Te$_3$~\cite{hor2010,fu2010,wangzw2016}. Recently, the QAH with higher Chern numbers has been realized in a multilayer of MTI~\cite{he2018}. Such experimental progress on the material growth and rich material choice of MTI and T$s$-SC makes the realization of the higher odd $N$ chiral TSC feasible. The transport of QTQ junctions of other higher $N$ chiral TSC will be studied in future work. 

\begin{acknowledgments}
\emph{Acknowledgments.} We thank Yang Feng, Xiao Feng, Tong Zhang and Ke He for helpful discussions. J.W. is supported by the Natural Science Foundation of China through Grant No.~11774065; the National Key Research Program of China under Grant No.~2016YFA0300703; the Natural Science Foundation of Shanghai under Grant No.~17ZR1442500; the National Thousand-Young-Talents Program; the Open Research Fund Program of the State Key Laboratory of Low-Dimensional Quantum Physics, through Contract No.~KF201606; and by Fudan University Initiative Scientific Research Program. B.L. is supported by the Princeton Center for Theoretical Science at Princeton University.
\end{acknowledgments}

\end{document}